\documentclass[pra, twocolumn, floatfix]{revtex4}
\usepackage{graphicx}
\usepackage{epstopdf}
\usepackage{amsmath, amsfonts, amssymb, bm,color}
\begin{document}
\title{Microwave multiphoton conversion via coherently driven permanent dipole systems}
\author{Alexandra \surname{M\^{i}rzac}}
\author{Sergiu \surname{Carlig}}
\author{Mihai A. \surname{Macovei}}
\email{mihai.macovei@ifa.md}
\affiliation{Institute of Applied Physics, Academiei str. 5, MD-2028 Chi\c{s}in\u{a}u, Moldova}
\date{\today}
\begin{abstract}
We investigate the multiphoton quantum dynamics of a leaking single-mode quantized cavity field coupled with a resonantly driven two-level system 
possessing permanent dipoles. The frequencies of the interacting subsystems are being considered very different, e.g., microwave ranges for the 
cavity and optical domains for the frequency of the two-level emitter, respectively. In this way, the emitter couples to the resonator mode via its 
diagonal dipole moments only. Furthermore, the generalized Rabi frequency resulting form the external coherent driving of the two-level subsystem 
is assumed as well different from the resonator's frequency or its multiples. As a consequence, this highly dispersive interaction regime is responsible 
for the cavity multiphoton quantum dynamics and photon conversion from optical to microwave ranges, respectively. 
\end{abstract}
\maketitle

\section{Introduction}
Frequency conversion processes where an input light beam can be converted at will into an output beam of a different frequency are very relevant 
nowadays due to various feasible quantum applications \cite{aps,aps1,aps2,aps3}. Among the first demonstrations of this effect is the experiment 
reported in \cite{exp1} promising developments of tunable sources of quantum light. From this reason, single-photon upconversion from a quantum 
dot preserving the quantum features was demonstrated in \cite{exp2}. Experimental demonstration of strong coupling between telecom (1550 nm) 
and visible (775 nm) optical modes on an aluminum nitride photonic chip was demonstrated as well, in Ref.~\cite{exp3}. Even bigger frequency 
differences can be generated. For instance, an experimental demonstration of converting a microwave field to an optical field via frequency 
mixing in a cloud of cold ${}^{87}$Rb atoms was reported in \cite{exp4}. Earlier theoretical studies have demonstrated frequency downconversion 
in pumped two-level systems with broken inversion symmetry \cite{theor1,theor2}. Furthermore, single- and multiphoton frequency conversion via 
ultra-strong coupling of a two-level emitter to two resonators was theoretically predicted in \cite{theor3}.  Although multiquanta processes are being 
investigated already for a long period of time, recently have attracted considerable attention as well. This is mainly due to potential application of 
these processes to quantum technologies related to quantum lithography \cite{qlit} or novel sources of light \cite{nsource}, etc. \cite{rew,ffilter,fedr}. 
Additionaly, optomechanically multiphonon induced transparency of x-rays via optical control was demonstrated in \cite{wen-te} while strongly 
correlated multiphonon emission in an acoustical cavity coupled to a driven two-level quantum dot was demonstrated in Ref.~\cite{nori}, respectively.

\begin{figure}[t]
\includegraphics[height=3.7cm]{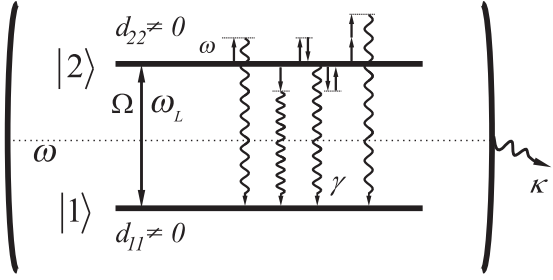}
\caption{\label{fig-0} 
The schematic of the model: A coherently pumped two-level emitter couples with a single-mode resonator of frequency $\omega$ via its non-zero 
diagonal dipoles, $d_{\alpha\alpha}$, with $\alpha \in \{1,2\}$. Here, $\Omega$ is the corresponding Rabi frequency due to the off-diagonal dipole 
moment $d_{21}$ whereas $\omega_{L}$, $\omega_{L} \gg \omega$, is the frequency of the resonantly applied external field. The emitter-resonator 
coupling strength is denoted by $g$, while $\kappa$ is the resonator's decay rate. Also, are sketched some processes which may occur, namely, 
emission/absorption of a cavity photon (or two, three, etc.) followed by the spontaneous decay $\gamma$.}
\end{figure}

However, most of the frequency conversion investigations refer to resonant processes. In this context, here, we shall demonstrate a photon 
conversion scheme involving non-resonant multiphoton effects, respectively. Actually, we investigate frequency downconversion processes via 
a resonantly laser-pumped two-level emitter possessing permanent diagonal dipoles, $d_{\alpha\alpha}\not=0$ with $\alpha \in \{1,2\}$, and 
embedded in a single-mode quantized resonator, see Fig.~(\ref{fig-0}). The frequency of the two-level emitter is assumed to be in the optical 
range and it is significantly different from the cavity frequency which may be in the microwave domain, for instance. Therefore, the two-level 
emitter naturally couples to the resonator through its permanent dipoles only. The cavity's frequency or its multiples differs as well from the 
generalized Rabi frequency arising due to resonant and coherent external driving of the two-level emitter. As a result, this highly dispersive 
interaction regime leads to multiphoton absorption-emission processes in the resonator mode mediated by the corresponding damping effects, 
i.e., emitter's spontaneous emission and the photon leaking through the cavity walls, respectively. We have obtained the corresponding cavity 
photon quantum dynamics in the steady state and demonstrated the feasibility to generate a certain multiphoton superposition state with high 
probability, and at different frequencies than that of the input external coherent pumping. The multiquanta nature of the final cavity state can 
be demonstrated via the second-order photon-photon correlation function. 

The advantage of our scheme consists in availability of its constituents, having $d_{22}\not=d_{11}$, such as asymmetrical two-level quantum 
dots \cite{twoa,mqd1,mqd2} and molecules \cite{kov,alt1,alt2}, or, equivalently, spin or quantum circuits \cite{alt3,alt4}, together with the 
technological progress towards their coupling to various resonators. As feasible applications of our results one may consider the possibility to 
couple distant real or artificial atoms having transition frequencies in the microwave domain via the multiphoton state generated by the developed
model here, see also \cite{dist_at}. Various entangled states  \cite{enth1,enth2,basharov,ficek} of distant emitters can be generated then. Another 
option, for instance, would be to investigate the quantum thermodynamic performances \cite{quth} of distant qubit systems interconnected through the 
microwave multiphoton field described here.

The article is organized as follows. In Sec. II we apply the developed analytical approach to the system of interest and describe it, while in Sec. III 
we analyze the obtained results. The summary is given in Sec. IV.

\section{Analytical framework}
The master equation describing the interaction of a two-level emitter, possessing permanent diagonal dipoles, with a classical coherent electromagnetic 
field of frequency $\omega_{L}$ as well as with a quantized single mode resonator of frequency $\omega$, with $\omega \ll \omega_{L}$ 
(see Fig.~\ref{fig-0}), and damped via the corresponding environmental reservoirs in the Born-Markov approximations \cite{agar,szu,gxl_b}, is:
\begin{eqnarray}
\frac{d}{dt}\rho(t) &+& \frac{i}{\hbar}[H,\rho] = -\frac{\gamma}{2}[S^{+},S^{-}\rho] - \frac{\kappa}{2}(1+\bar n)[b^{\dagger},b\rho] 
\nonumber \\
&-& \frac{\kappa}{2}\bar n[b,b^{\dagger}\rho] + H.c..
\label{meq}
\end{eqnarray}
Here, $\gamma$ is the single-emitter spontaneous decay rate, whereas $\kappa$ is the corresponding boson-mode's leaking rate with 
$\bar n$ = $\bigl[\exp[\hbar\omega/(k_{B}T)]-1\bigr]^{-1}$ being the mean resonator’s photon number due to the environmental 
thermostat at temperature $T$, and $k_{B}$ is the Boltzmann constant. The two-level system may have the transition frequency in 
the optical domain, whereas the single-mode cavity frequency may lay in the microwave range, respectively. The wavevector of the 
coherent applied field is perpendicular to the cavity axis.  In the Eq.~(\ref{meq}), the bare-state emitter's operators 
$S^{+}=|2\rangle \langle 1|$ and $S^{-}=[S^{+}]^{\dagger}$ obey the commutation relations for su(2) algebra, namely, 
$[S^{+},S^{-}]=2S_{z}$ and $[S_{z},S^{\pm}]=\pm S^{\pm}$, where $S_{z}=(|2\rangle \langle 2| - |1\rangle \langle 1|)/2$ is 
the bare-state inversion operator. $|2\rangle$ and $|1\rangle$ are the excited and ground state of the emitter, respectively, while 
$b^{\dagger}$ and $b$ are the creation and the annihilation operator of the electromagnetic field (EMF) in the resonator, and 
satisfy the standard bosonic commutation relations, i.e., $[b,b^{\dagger}]=1$, and $[b,b]=[b^{\dagger},b^{\dagger}]=0$. 
The Hamiltonian characterizing the respective coherent evolution of the considered compound system is (see Appendix A):
\begin{eqnarray}
H = \hbar \omega b^{\dagger}b + \hbar \Delta S_{z} - \hbar\Omega(S^{+} + S^{-}) + \hbar g S_{z}(b^{\dagger}+ b).
\label{H}
\end{eqnarray}
In the Hamiltonian (\ref{H}), the first two components describe the free energies of the cavity electromagnetic field and the two-level emitter, respectively, 
with $\Delta=\omega_{21}-\omega_{L}$ being the detuning of the emitter transition frequency $\omega_{21}$ from the laser one. The last two terms 
depict, respectively, the laser interaction with the two-level system and the emitter-cavity interaction. $\Omega$ and $g$ are the corresponding coupling 
strengths. Note at this stage that while the Rabi frequency $\Omega$ is proportional to the off-diagonal dipole moment $d_{21}$, the emitter-cavity 
coupling is proportional to the diagonal dipole moments, i.e. $g\propto (d_{22}-d_{11})$. The interaction of the external coherent electromagnetic field 
with permanent dipoles is omitted here as being rapidly oscillating. From the same reason, the emitter-cavity interaction described by the usual 
Jaynes-Cummings Hamiltonian, proportional to $d_{21}$, is neglected as well here, see Appendix A.
 
In what follows, we perform a spin rotation \cite{enaki,saiko,mekPRA2020}, 
\begin{equation}
U(\chi)=\exp\bigl[2i\chi S_{y}\bigr], 
\label{spR}
\end{equation}
where $S_{y}=(S^{+} - S^{-})/2i$ and $2\chi=\arctan{[2\Omega/\bar\Delta]}$ with $\bar \Delta = \Delta + g(b^{\dagger} + b)$, diagonalizing the 
last three terms of the Hamiltonian (\ref{H}). This action will lead to new quasi-spin operators, i.e. $R_{z}$ and $R^{\pm}$, defined via the old emitter's 
operators in the following way
\begin{eqnarray}
R_{z} &=&S_{z}\cos{2\chi} - (S^{+} + S^{-})\sin{2\chi}/2, \nonumber \\
R^{+} &=&S^{+}\cos^{2}{\chi} - S^{-}\sin^{2}{\chi} + S_{z}\sin{2\chi}, \nonumber \\
R^{-} &=& [R^{+}]^{\dagger}. 
\label{ns}
\end{eqnarray}
The new emitter operators $R^{+}=|\bar 2\rangle \langle \bar 1|$, $R^{-}=|\bar 1\rangle \langle \bar 2|$ and $R_{z}=(|\bar 2\rangle \langle \bar 2| 
- |\bar 1\rangle \langle \bar 1| )/2$, describing the transitions and populations among the dressed-states $\{|\bar 2\rangle, |\bar 1\rangle\}$, will 
obey the commutation relations: $[R^{+},R^{-}]=2R_{z}$ and $[R_{z},R^{\pm}]=\pm R^{\pm}$, similarly to the old-basis ones. Respectively, 
the Hamiltonian (\ref{H}) transforms to:
\begin{eqnarray}
\bar H = \hbar \omega b^{\dagger}b + 2\hbar \bar\Omega R_{z},
\label{HH}
\end{eqnarray}
where the operator $\bar \Omega = (\bar \Delta^{2}/4 + \Omega^{2})^{1/2}$, whereas
\begin{equation}
b = \bar b - i\eta S_{y} \sum^{\infty}_{k=0}\frac{\eta^{k}}{k!}(\bar b^{\dagger} 
+ \bar b)^{k}\frac{\partial^{k}}{\partial \xi^{k}}\frac{1}{1+\xi^{2}},
\label{bb}
\end{equation}
with $b^{\dagger} = [b]^{\dagger}$, $\bar b = U{b}U^{-1}$, $\bar b^{\dagger} = [\bar b]^{\dagger}$, and
\begin{equation*}
\eta = \frac{g}{2\Omega}, ~~{\rm and} ~~\xi = \frac{\Delta}{2\Omega}.
\end{equation*}
Now the expressions  (\ref{ns})-(\ref{bb}) have to be introduced in the master equation (\ref{meq}) and the final equation will be somehow cumbersome. 
It can be simplified if we perform the secular approximation, i.e., neglect all terms from the master equation oscillating at the generalized Rabi frequency 
$2\Omega_{0}$, $\Omega_{0} = \Omega\sqrt{1+\xi^{2}}$, and higher one. This is justified if $2\Omega_{0} \gg \{g,\gamma\}$ - the situation considered 
here. 

Thus, in the following, we expand the generalized Rabi frequency $\bar \Omega$ in the Taylor series using the small parameter $\eta$, namely, 
\begin{equation*}
\bar \Omega = \Omega_{0}\biggl\{1 + \frac{\xi \hat \eta}{1+\xi^{2}} + \frac{\hat \eta^{2}}{2(1+\xi^{2})^{2}} 
- \frac{\xi \hat \eta^{3}}{2(1+\xi^{2})^{3}} + \cdots \biggr\},
\end{equation*}
where $\hat \eta = \eta(\bar b^{\dagger} + \bar b)$. Then perform a unitary transformation $U(t)=\exp[2i\Omega_{0}R_{z}t]$ in the whole master 
equation and neglect terms oscillating at the Rabi frequency $2\Omega_{0}$ or higher. Afterwards, perform the operation $\rho_{\bar \alpha \bar \alpha} = 
\langle \bar \alpha|\rho|\bar \alpha\rangle$, $\alpha \in \{1,2\}$, and one can arrive then at the following master equation describing the cavity degrees 
of freedom only: 
\begin{eqnarray}
\frac{d}{dt}\bar \rho(t) &+& \frac{i}{\hbar}[\bar H,\bar \rho] - \frac{\gamma}{4}\{\cos2\chi\bar \rho\cos2\chi + \sin2\chi\bar \rho\sin2\chi \nonumber \\
&-& \bar \rho\} = - \frac{\kappa}{2}(1+\bar n)[\bar b^{\dagger},\bar b\bar \rho] - \frac{\kappa}{2}\bar n[\bar b,\bar b^{\dagger}\bar \rho] 
\nonumber \\
&-&\frac{\kappa\eta^{2}}{8}(1+2\bar n)\sum^{\infty}_{k_{1},k_{2}=0}f_{k_{1}}(\eta,\xi)f_{k_{2}}(\eta,\xi)
\nonumber \\
&\times& [(\bar b^{\dagger}+ \bar b)^{k_{1}},(\bar b^{\dagger}+ \bar b)^{k_{2}}\bar \rho] + H.c.,
\label{mqq}
\end{eqnarray}
where $\bar \rho=\rho_{\bar 1\bar 1} + \rho_{\bar 2 \bar 2}$.
Here, 
\begin{eqnarray*}
f_{k}(\eta,\xi) &=& \frac{\eta^{k}}{k!}\frac{\partial^{k}}{\partial \xi^{k}}\frac{1}{1 + \xi^{2}}, \nonumber \\
\sin2\chi &=& \frac{\Omega}{\bar \Omega} = \sum^{\infty}_{k=0}\frac{\eta^{k}(\bar b + \bar b^{\dagger})^{k}}{k!}
\frac{\partial^{k}}{\partial \xi^{k}}\frac{1}{\sqrt{1+\xi^{2}}}, \nonumber \\
\cos2\chi &=& \frac{\bar \Delta/2}{\bar \Omega} = \sum^{\infty}_{k=0}\frac{\eta^{k}(\bar b + \bar b^{\dagger})^{k}}{k!}
\frac{\partial^{k}}{\partial \xi^{k}}\frac{\xi}{\sqrt{1+\xi^{2}}}.
\end{eqnarray*}
Already at this stage one can recognize the multiphoton nature of the cavity electromagnetic field quantum dynamics. Particularly in Eq.~(\ref{mqq}), 
the term proportional to $\gamma$ describes the resonator's multiphoton dynamics accompanied by the spontaneous decay, whereas the components 
proportional to $\kappa$ characterize the same processes but followed by the cavity decay, respectively, see also Fig.~(\ref{fig-0}). 

Using the bosonic operator identity \cite{ref_binom}
\begin{eqnarray*}
(A+ B)^{n} &=& \sum^{n}_{k'}\frac{n!}{k'!(\frac{n-k'}{2})!}\biggl(-\frac{C}{2}\biggr)^{\frac{n-k'}{2}}\sum^{k'}_{r=0}\frac{k'!}{r!(k'-r)!}
\nonumber \\
&\times&A^{r}B^{k'-r},
\end{eqnarray*}
where $[A,B]=C$ and $[A,C]=[B,C]=0$, whereas $k'$ is odd for an odd $n$ and even for an even $n$ (if, for instance, $n=4$, then $k'=\{0,2,4\}$, 
while if $n=5$, then $k'=\{1,3,5\}$, whereas $r=0,1,2,\cdots, k'$), one can reduce the master equation (\ref{mqq}) to a time-independent equation 
if one further performs a unitary transformation $V(t)=\exp[i\omega\bar b^{\dagger}\bar b t]$ and neglects all the terms rotating at frequency 
$\omega$ and higher. This would also result in avoiding any resonances in the system, i.e., $2\Omega_{0} - s\omega \not=0$, $s \in \{1,2,\cdots\}$. 
As a consequence, one can obtain a diagonal equation for $P_{n} = \langle n|\bar \rho|n\rangle$, with $|n\rangle$ being the Fock state and 
$n \in \{0,1,2,\cdots \}$, describing the cavity multiphoton quantum dynamics, in the presence of corresponding damping effects, which is computed 
then numerically here. Notice that the coherent part of the master equation (\ref{mqq}), i.e. $[\bar H,\bar \rho]$, does not contribute to the final 
expression for the photon distribution function $P_{n}$. The reason is that after the performed approximations the Hamiltonian $\bar H$ would 
contain photonic correlators such that $\langle n|[\bar H,\bar \rho]|n\rangle$=$\bar H_{n}P_{n}-P_{n}\bar H_{n}=0$.

Thus, the cavity photon dynamics has a multiphoton behavior because of the highly dispersive (non-resonant) nature of the interaction among the 
asymmetrical two-level emitter and cavity field mode. This way, one obtains an output multiphoton flux of microwave photons, although the two-level 
system is coherently pumped at a different frequency, i.e. with optical photons.

\section{Results and discussion}
\begin{figure}[t]
\includegraphics[width = 4.29cm]{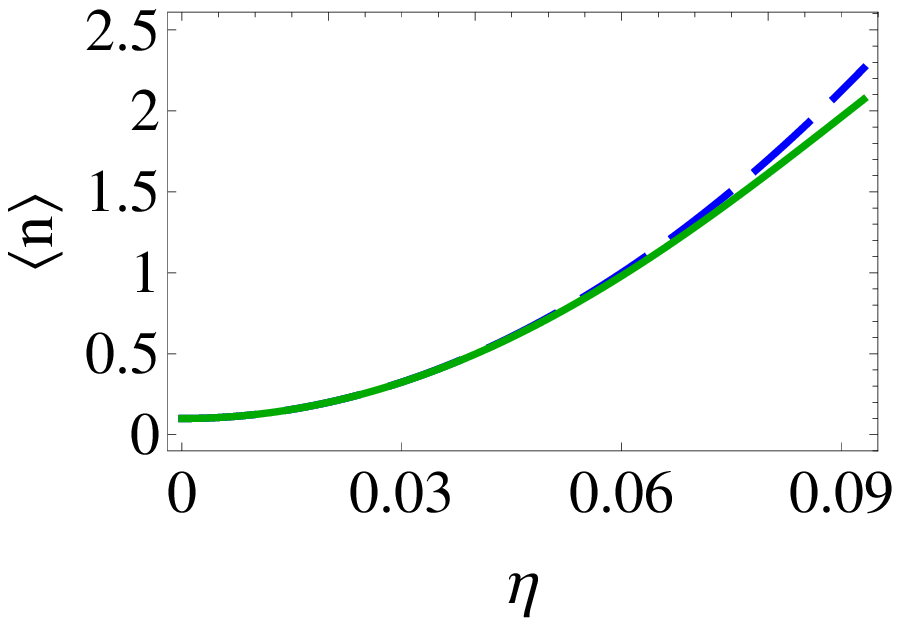}
\hspace{-0.2cm}
\includegraphics[width = 4.35cm]{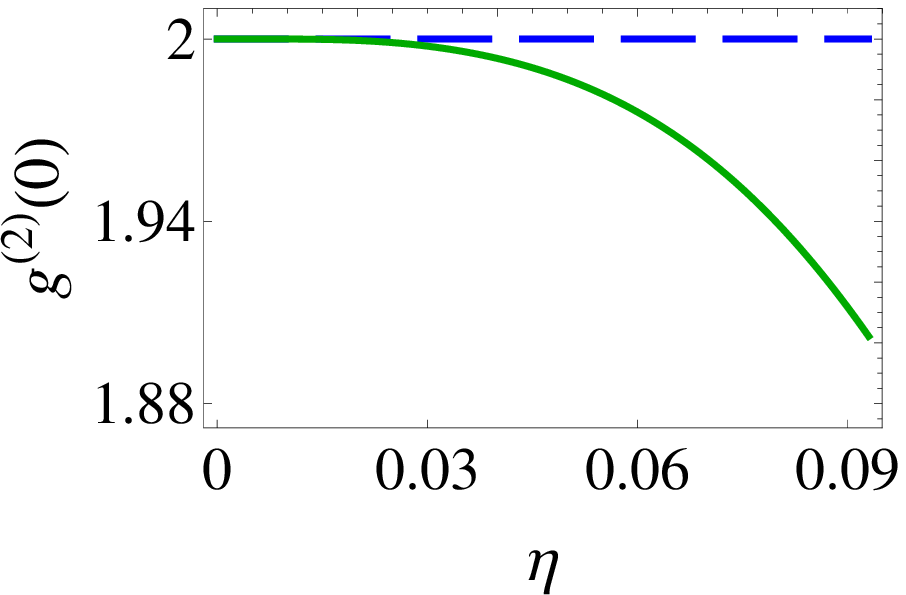}
\begin{picture}(0,0)
\put(-97,75){(a)}
\put(30,75){(b)}
\end{picture}
\caption{\label{fig-1} 
(a) The steady-state mean cavity photon number $\langle n\rangle \equiv \langle \bar b^{\dagger}\bar b\rangle$ as well as (b) its second-order 
correlation function $g^{(2)}(0)$ as a function of $\eta = g/(2\Omega)$. The blue long-dashed lines are plotted for single-photon processes, N=1, 
while the solid green ones for two-photon processes, N=2, respectively. Here, $\bar n=10^{-1}$, $\kappa/\gamma=10^{-3}$ and $\xi=0$.}
\end{figure}

In the following, we shall describe the cavity multiphoton quantum dynamics based on the Eq.~(\ref{mqq}). Particularly, for single-photon non-resonant 
processes one can obtain the following equation for the photon distribution function, see Appendix B: 
\begin{eqnarray}
\frac{d}{dt}P_{n}(t) = -P^{(1)}_{n}, \label{pnn1}
\end{eqnarray}
where
\begin{eqnarray*}
 P^{(1)}_{n}&=&\biggl\{\kappa(1+\bar n)+ \frac{\gamma\eta^{2}}{4(1+\xi^{2})^{2}}\biggr\}\biggl ( nP_{n}- (n+1)P_{n+1} \biggr)\nonumber \\
&+&\biggl\{\kappa\bar n+ \frac{\gamma\eta^{2}}{4(1+\xi^{2})^{2}}\biggr\}\biggl ((n+1)P_{n} - nP_{n-1} \biggr).
\end{eqnarray*}
The first line of the above expression for $P^{(1)}_{n}$ describes the photon generation processes, i.e., photons that leave the cavity. 
The second line corresponds to processes describing photons pumping the cavity mode due to the environmental thermostat and 
non-resonant external driving, respectively. One can observe that both processes are influenced by the resonant laser pumping of 
the two-level emitter possessing permanent dipoles. As a consequence, the stationary mean-photon number in the resonator mode is,
see Appendix B:
\begin{eqnarray}
\langle \bar b^{\dagger}\bar b\rangle= \bar n + \frac{\gamma \eta^{2}}{4\kappa(1+\xi^{2})^{2}}, 
\label{bpb1}
\end{eqnarray}
whereas its second-order photon-photon correlation function is $g^{(2)}(0)=2$, see the blue long-dashed curves in Fig.~(\ref{fig-1}). Respectively, 
for two-photon non-resonant processes one has:
\begin{eqnarray}
\frac{d}{dt}P_{n}(t) = - P^{(2)}_{n}, \label{pnn2}
\end{eqnarray}
where, see Appendix B,
\begin{eqnarray*}
&{}&P^{(2)}_{n} =  P^{(1)}_{n} - \frac{3\gamma(1-2\xi^{2})\eta^{4}}{4(1+\xi^{2})^{4}} \nonumber \\
&\times& \biggl((1+n)^{2}(P_{n}-P_{n+1})+ n^{2}(P_{n}-P_{n-1})\biggr) \nonumber \\
&+& \frac{\gamma(1+4\xi^{2})\eta^{4}}{16(1+\xi^{2})^{4}}\biggl(n(n-1)P_{n} - (n+1)(n+2)P_{n+2}\biggr) \nonumber \\
&+& \frac{\gamma(1+4\xi^{2})\eta^{4}}{16(1+\xi^{2})^{4}}\biggl((n+1)(n+2)P_{n}-n(n-1)P_{n-2}\biggr),
\end{eqnarray*}
where smaller contributions, proportional to $\kappa\eta^{4}$, were neglected since we have considered that $\kappa/\gamma \ll 1$. Here, the first 
two lines of the expression for $P^{(2)}_{n}$ describe the photon depopulation and population of the cavity mode due to single-photon processes. 
Notice that the single-photon effects are influenced by the second-order one, see the second term proportional to $\eta^{4}$ in the first line of 
$P^{(2)}_{n}$. The last two lines of the same expression consider the resonator photon depopulation and population effects via two-photon 
processes, respectively. Thus, Eq.~(\ref{pnn2}) describes photon processes where single-photon and two-photon effects coexist simultaneously. 
As we will see latter, the mean-photon number in the cavity mode and its second-order photon-photon correlation functions change accordingly. 
Similarly, additional $N-photon$ non-resonant processes with $N \in \{3,4,\cdots\}$ can be incorporated by restricting the equation Eq.~(\ref{mqq}) 
to terms up to $\eta^{2N}$, see Appendix B. 

In order to solve the infinite system of equations for $P_{n}$ (see e.g. Eq. \ref{pnn2} for two-photon processes), we truncate it at a certain maximum 
value $n=n_{max}$ so that a further increase of its value, i.e. $n_{max}$, does not modify the obtained results if other involved parameters are being 
fixed. Thus, generally the resonator's steady-state mean quanta number can be expressed as:
\begin{eqnarray}
\langle \bar b^{\dagger} \bar b \rangle = \sum^{n_{max}}_{n=0}nP_{n}, \label{bpb}
\end{eqnarray}
with 
\begin{eqnarray}
\sum^{n_{max}}_{n=0}P_{n}=1. \label{nrm}
\end{eqnarray}
Respectively, the second-order photon-photon correlation function is defined in the usual way \cite{szu,glauber}, 
namely,
\begin{figure}[t]
\includegraphics[width = 4.29cm]{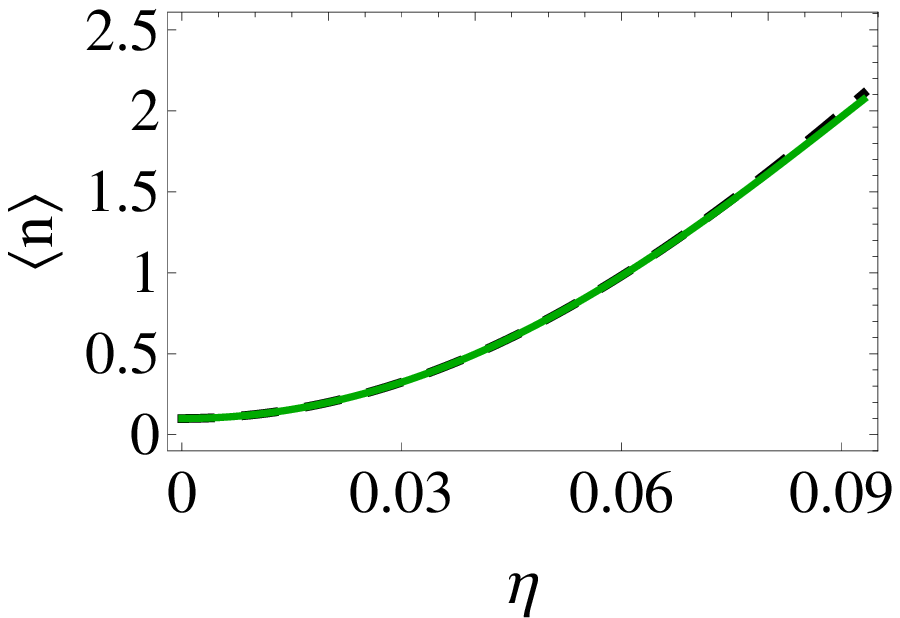}
\hspace{-0.2cm}
\includegraphics[width = 4.35cm]{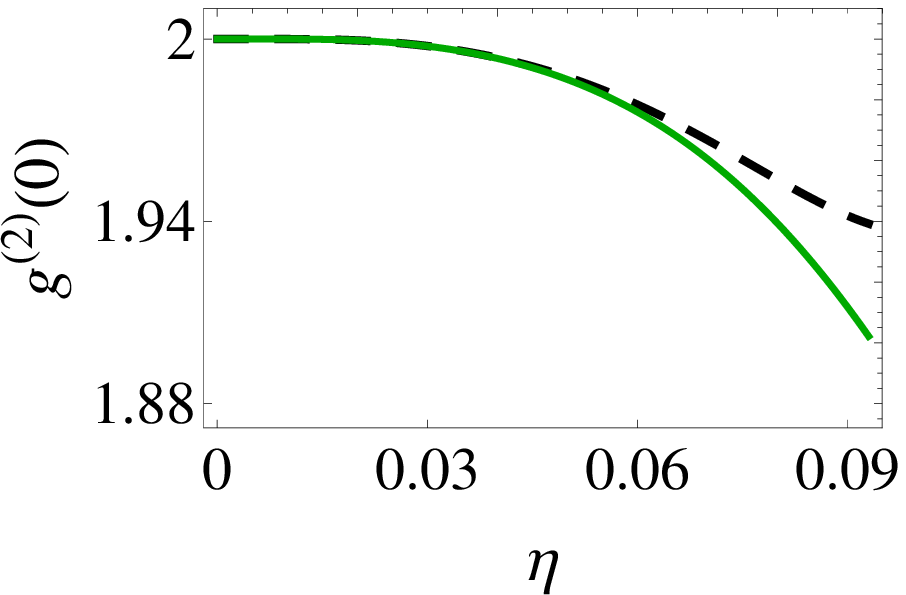}
\begin{picture}(0,0)
\put(-97,75){(a)}
\put(30,75){(b)}
\end{picture}
\caption{\label{fig-2} 
(a) The steady-state mean cavity photon number $\langle n\rangle \equiv \langle \bar b^{\dagger}\bar b\rangle$ as well as (b) its second-order 
correlation function $g^{(2)}(0)$ as a function of $\eta = g/(2\Omega)$. The solid green curves are plotted for two-photon processes, N=2, while 
the short-dashed black ones for three-photon processes, N=3, respectively. Other parameters are as in Fig.~(\ref{fig-1}).}
\end{figure}
\begin{eqnarray}
g^{(2)}(0) &=& \frac{\langle \bar b^{\dagger 2}\bar b^{2}\rangle}{\langle \bar b^{\dagger}\bar b\rangle^{2}} \nonumber \\
&=& \bigl(1/\langle \bar b^{\dagger}\bar b\rangle^{2}\bigr)\sum^{n_{max}}_{n=0}n(n-1)P_{n}. 
\label{gdoi}
\end{eqnarray}
Note here that we need to evaluate the cavity field correlators, i.e. $\langle b^{\dagger}b\rangle$ etc., using Eq.~(\ref{bb}) 
first. That is, one expresses $\langle b^{\dagger}b\rangle$ via $\langle \bar b^{\dagger}\bar b\rangle$ and calculate the latter 
correlator using the above developed approach. From Eq.~(\ref{bb}) and within the performed approximations, one can observe 
however, that $\langle b^{\dagger}b\rangle$=$\langle \bar b^{\dagger}\bar b\rangle$ + $o(\eta^{4})$. Therefore, for $\eta \ll 1$, 
as it is the case considered here, we have $\langle b^{\dagger}b\rangle \approx \langle \bar b^{\dagger}\bar b\rangle$, and one 
can surely use the field operators $\{\bar b,\bar b^{\dagger}\}$ to calculate the cavity mean-photon number and its second-order 
photon-photon correlations via expressions (\ref{bpb}-\ref{gdoi}).

Thus, Figure (\ref{fig-1}) shows the steady-state mean photon numbers and their second-order photon-photon correlation 
functions for single-photon and two-photon processes plotted with the help of Eq.~(\ref{pnn1}) and Eq.~(\ref{pnn2}). One can 
observe here that these quantities differ from each other for single- and two-photon effects, respectively. For the sake of 
comparison, Figure (\ref{fig-2}) depicts similar things for two- and three-photon effects, correspondingly. Here, it is easy to see 
that the mean-photon numbers almost overlap for the two cases considered, whereas their second-order correlation functions 
distinguish from each other. One can proceed in the same vein with higher order photon processes. However, for identically 
considered parameters, their probabilities are small and the mean photon numbers are basically the same as indicated in 
Fig.~\ref{fig-2}(a). On the other side, the photon statistics exhibits quasi-thermal features as $\eta$ increases with other 
parameters being fixed. Concluding this part, more probable are processes with single-, two- and three-photons, respectively, 
if other involved parameters are being fixed, whereas the final cavity steady state is a quantum incoherent superposition of all 
those photons. Importantly, values different from $2$ for $g^{(2)}(0)$, occurring naturally for higher values of $\eta's$ with 
$\eta <1$, ensure the creation of this final cavity state which differs from a usual thermal state. Note that generally the 
environmental thermal mean-photon number will add linearly to the final photon flux (see, e.g., Eq.~\ref{bpb1} for single-photon 
processes) so that an increase in the environmental temperature will lead to more output photons for the considered parameter 
ranges.
\begin{figure}[t]
\includegraphics[width=7.5cm]{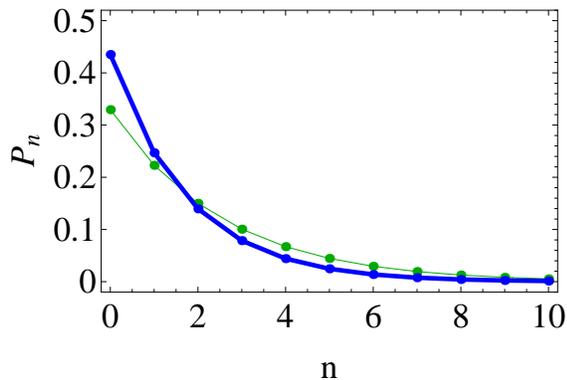}
\caption{\label{fig-3} 
The cavity photon  distribution function $P_{n}$ in the steady state. The green thin curve is plotted for $\eta=0.09$, 
while the thick blue one for $\eta=0.07$, respectively. Other parameters are as in Fig.~(\ref{fig-1}).}
\end{figure}

Additionally, Figure (\ref{fig-3}) shows the photon distribution function $P_{n}=\langle n|\bar \rho |n\rangle$ for the same parameters as 
taken in Figs.~(\ref{fig-1}) and (\ref{fig-2}), however, for five-photon processes, i.e. $N=5$. One can observe here that larger ratios of 
$\eta=g/(2\Omega)$, with $\eta < 1$,  lead to population of higher photon states, compare the thin green and thick blue curves plotted for 
$\eta=0.09$ and $\eta=0.07$, respectively, facilitating the generation of multiphoton states when $\kappa/\gamma \ll 1$. Correspondingly, 
$P_{n}$ is small for larger $n$ and smaller $\eta$, while $\eta < 1$, assuring convergence of the results based on Eq.~(\ref{mqq}). One can 
observe that the probability of a two-photon state, that is $n=2$, is almost the same for $\eta=0.07$ and $\eta=0.09$, respectively, and it 
is higher than $0.1$. Thus a multiphoton superposition state around $n=2$ is generated when other parameters are being fixed. Furthermore, 
the same results, shown in the above figures will persist for moderate detunings, i.e., would not change significantly 
if $\xi \ll 1$. 

Concluding here, the presence of diagonal dipole moments, in a resonance coherently pumped two-level system, makes possible the coupling to 
the resonator mode at a completely different frequency than the input one which drives the two-state quantum emitter, and cavity multiphoton 
state generation, respectively. Furthermore, the developed approach applies equally to a driven two-level quantum dot embedded in an acoustical 
phonon cavity see, e.g. \cite{nori,knor,cm,erem}. It can be generalized as well to an ensemble of two-level emitters \cite{agar} having permanent 
dipoles and embedded in a microwave resonator. Finally notice that the results shown in Figs.~(\ref{fig-1}-\ref{fig-3}) can be obtained directly 
by a full simulation of the master equation (\ref{meq}). However, in this case, one can not extract information about incoherent multiphoton processes 
that originate the final cavity steady state.

\section{Summary}
We have investigated the possibility to convert photons from, e.g. optical to microwave domains, via a resonantly pumped asymmetrical two-level 
quantum emitter embedded in a quantized single-mode resonator. The corresponding damping effects due to emitter's spontaneous emission and 
cavity's photon leakage are taking into account as well. The transition frequency of the two-level system differs significantly from the cavity's one, 
namely, it can lay in the optical range while the resonator's frequency in the microwave domain, respectively. Therefore, the two-state quantum 
emitter couples to the cavity mode through its diagonal dipole moments. As well, the cavity's frequency is considering being far off-resonance 
from the generalized Rabi frequency resulting from the coherent driving of the two-level system via its non-diagonal dipole. In these circumstances, 
multiphoton absorption-emission processes are proper to the cavity quantum dynamics. We have demonstrated the cavity's multiphoton 
characteristics and showed the feasibility for a certain output multiphoton superposition state generation. The photon statistics exhibits quasi-thermal 
photon statistics as the pumping parameter $\eta$ is increased from zero. Actually, values different from $2$ for the second-order photon-photon 
correlation function $g^{(2)}(0)$ ensure the creation of the cavity multiphoton superposition state. Finally, as a 
concrete system, where the approach developed here can apply, can serve asymmetrical two-level quantum dots coupled to microwave 
resonators as well as polar biomolecules, spin or quantum circuit systems, respectively \cite{twoa,mqd1,mqd2,kov,alt1,alt2,alt3,alt4}. In principle, 
coupling to terahertz or even higher-frequency resonators will allow photon conversion in these photon ranges too. As well, this analytical 
approach can be used to study non-resonant multiphonon quantum dynamics when a pumped two-level quantum dot interacts with an acoustical 
phonon resonator, respectively \cite{nori,knor,cm,erem}. Finally, it can be generalized to an ensemble of two-level emitters \cite{agar} 
having permanent dipoles.

\acknowledgments
We acknowledge the financial support by the Moldavian National Agency for Research and Development, grant No. 20.80009.5007.07. 
Also, A.M. is grateful to the financial support from National Scholarship of World Federation of Scientists in Moldova.

\appendix
\section{The system's Hamiltonian}
Here we present details on how one arrives at the system's Hamiltonian given by Eq.~(\ref{H}). The complete Hamiltonian 
describing the interaction of a two-level emitter possessing permanent dipoles with an external resonant coherent field as well as 
with a single-mode resonator, in the dipole and rotating wave approximations,  is:
\begin{eqnarray}
H&=&\hbar\omega b^{\dagger}b + \hbar\omega_{21}S_{z} - \hbar\Omega\bigl(S^{+}e^{-i\omega_{L}t} + S^{-}e^{i\omega_{L}t} \bigr) 
\nonumber \\
&+&\hbar g_{0}\bigl(d_{22}S_{22} + d_{11}S_{11}\bigr)\bigl(b^{\dagger}+ b\bigr) + \hbar\bar g_{0}\bigl(S^{+}+ S^{-}\bigr)  \nonumber \\ 
&\times&\bigl(b^{\dagger}+ b\bigr) - E_{L}\bigl(d_{22}S_{22} + d_{11}S_{11}\bigr)\cos(\omega_{L}t). \label{HT}
\end{eqnarray}
Here the first two terms describe the free energies of the resonator and the two-level subsystem. The third and the sixth terms account for the 
interaction of the external laser field with the two-level emitter through its off-diagonal dipole moments $d_{21}$, $d_{21} = d_{12}$, as 
well as the diagonal dipole moments $d_{22}$ and $d_{11}$, respectively. Correspondingly, the fourth and the fifth components describe the 
interactions of the cavity mode with the two-level emitter via diagonal and off-diagonal dipole moments. Here, $E_{L}$ is the amplitude of the 
external driving field, while $g_{0}=\sqrt{2\pi\omega/\hbar V}$ where $V$ is the quantization volume, and $\bar g_{0}=g_{0}d_{21}$. 
$S_{\alpha\alpha}$, $\{\alpha = 1,2\}$, are the population operators, respectively. All other parameters and operators are described in Section II. 

After performing a unitary transformation $\bar U(t)=\exp{(i\omega_{L}S_{z}t)}$ one can observe that the fifth Hamiltonian's term is a 
rapidly oscillating one since $\omega_{L}$ is bigger than the corresponding coupling strength, i.e., 
$\omega_{L} \gg \bar g_{0}$ and $\omega_{L} \gg \omega$. As well, the last component of the Hamiltonian (\ref{HT}) 
can be neglected from the same reason because $\omega_{L} \gg \{E_{L}d_{22}/\hbar, E_{L} d_{11}/\hbar\}$ for moderate assumed 
external pumping strengths. Thus, one has then the following Hamiltonian
\begin{eqnarray}
H&=&\hbar\omega b^{\dagger}b + \hbar\Delta S_{z} - \hbar\Omega\bigl(S^{+} + S^{-} \bigr)  + \hbar g S_{z}\bigl(b^{\dagger}+ b\bigr) \nonumber \\
&+& \hbar g_{0}(d_{11}+d_{22})(b^{\dagger}+ b\bigr)/2, \label{HT1}
\end{eqnarray}
where $g=g_{0}(d_{22}-d_{11})$, and we have used also the relations $S_{22}=1/2 + S_{z}$, and $S_{11}=1/2 - S_{z}$. Further, performing 
a unitary transformation $V=\exp{(\zeta b -\zeta^{\ast}b^{\dagger})}$, with $\zeta=g_{0}(d_{11}+d_{22})/[2(\omega+i\kappa/2)]$, in the whole 
master equation (\ref{meq}), containing the Hamiltonian (\ref{HT1}), one arrives at the same form of the master equation with, however, the 
Hamiltonian (\ref{H}), and where $\Delta \equiv \Delta - g^{2}_{0}(d^{2}_{22}-d^{2}_{11})/\omega$, when $\omega \gg \kappa$. 
The last term from the detuning's expression can be used to redefine the emitter's frequency, i.e., $\omega_{21} \equiv \omega_{21} -  
g^{2}_{0}(d^{2}_{22}-d^{2}_{11})/\omega$, so one finally has $\Delta=\omega_{21}-\omega_{L}$.

Now, if we make a unitary transformation in the Hamiltonian (\ref{H}), $\bar V(t)=\exp{(i\omega b^{\dagger}bt)}$, then it transforms as:
\begin{eqnarray}
H= \hbar\Delta S_{z} - \hbar\Omega\bigl(S^{+} + S^{-} \bigr)  + \hbar g S_{z}\bigl(b^{\dagger}e^{i\omega t}+ be^{-i\omega t}\bigr). \nonumber \\
\label{HR}
\end{eqnarray}
If one avoids any resonances in the system with respect to the resonator's frequency or its multiples, as it is the case here, then the last term in the 
above Hamiltonian is a rapidly oscillating one, if $\omega$ is significantly larger than $g$, and may be neglected. Section II develops an approach 
where the contribution of this term is perturbatively calculated for moderately intense externally applied fields and appropriate parameters ranges,
i.e. $\omega > 2\Omega \gg \{g,\gamma,\kappa\}$, respectively.

\section{The master equation (\ref{mqq}) containing terms up to $\eta^{4}$}
Here, we shall emphasize some processes occurring in our setup in more details, namely, the single- and two-photon effects. 
Let's write down the time-independent damping part of the master equation (\ref{mqq}), taking into account expansion terms 
up to $\eta^{4}$, namely,
\begin{widetext}
\begin{eqnarray}
\frac{d}{dt}\bar \rho &=& -\frac{\gamma\eta^{2}}{8(1+\xi^{2})^{2}}\biggl\{\bigl[\bar b,\bar b^{\dagger}\bar \rho\bigr] 
+ \bigl[\bar b^{\dagger},\bar b\bar \rho\bigr]\biggr\} - \frac{\gamma\eta^{4}(1+4\xi^{2})}{32(1+\xi^{2})^{4}}\biggl\{\bigl[(\bar b\bar b^{\dagger} 
+ \bar b^{\dagger} \bar b),(\bar b\bar b^{\dagger} + \bar b^{\dagger} \bar b)\bar \rho\bigr] + \bigl[\bar b^{2},\bar b^{\dagger 2}\bar \rho \bigr] 
+ \bigl[\bar b^{\dagger 2},\bar b^{2}\bar \rho\bigr]\biggr\} \nonumber \\
&+& \frac{3\gamma\eta^{4}(1-2\xi^{2})}{8(1+\xi^{2})^{4}}\biggl \{\bigl[\bar b^{\dagger}(1+\bar b^{\dagger}\bar b),\bar b\bar \rho \bigr]  
+ \bigl[(1+\bar b^{\dagger}\bar b)\bar b,\bar b^{\dagger}\bar \rho\bigr] \biggr\} - \frac{\kappa}{2}(1+\bar n)\bigl[\bar b^{\dagger},\bar b\bar \rho\bigr] 
- \frac{\kappa}{2}\bar n\bigl[\bar b,\bar b^{\dagger}\bar \rho \bigr] + H.c.,
\label{two_B}
\end{eqnarray}
\end{widetext}
where smaller contributions, proportional to $\kappa\eta^{4}$, were neglected since we have considered that $\kappa/\gamma \ll 1$.

One can observe that terms proportional to $\eta^{2}$ describe single-photon processes, that is, the photon number in the distribution 
function $P_{n}$ ($P_{n} = \langle n|\bar \rho|n\rangle$ with $n \in \{0,1,2,\cdots \}$) will change by $\pm 1$, i.e. $P_{n \pm 1}$, see 
also Eq.~(\ref{pnn1}). Respectively, the terms proportional to $\eta^{4}$ account for two-photon effects. For instance, the last two 
commutators from the second term of Eq.~(\ref{two_B}) will modify the photon number in the distribution function $P_{n}$  by 
$\pm 2$, i.e. $P_{n \pm 2}$, see also Eq.~(\ref{pnn2}) and Fig.~(\ref{fig-0}). Concluding this part, one can generalize that terms 
proportional to $\eta^{2N}$, in the master equation (\ref{mqq}), account for $N$-photon processes, respectively. 

From Eq.~(\ref{two_B}) one can easily arrive at Eq.~(\ref{pnn2}). Setting then $\eta^{4}\to 0$, we obtain the Eq.~(\ref{pnn1}).
The steady-state solution of Eq.~(\ref{pnn1}), accounting for single-photon processes only, can be expressed as:
\begin{eqnarray}
P_{n}=Z^{-1}e^{-\alpha n}, \label{pnB}
\end{eqnarray}
where the normalization $Z$ is determined by the requirement $\sum^{\infty}_{n=0}P_{n}=1$, that is $Z=\sum^{\infty}_{n=0}e^{-\alpha n}$, 
whereas $\alpha = {\rm ln}\beta$ and $\beta = \kappa_{1}/\kappa_{2}$ with $\kappa_{1}=\kappa(1+\bar n)$+$\gamma\eta^{2}/[4(1+\xi^{2})^{2}]$, 
and $\kappa_{2}=\kappa\bar n$+$\gamma\eta^{2}/[4(1+\xi^{2})^{2}]$. The mean-photon number is determined via
\begin{eqnarray}
\langle \bar b^{\dagger}\bar b\rangle = \sum^{\infty}_{n=0}nP_{n} = \frac{1}{\beta -1} 
= \bar n + \frac{\gamma\eta^{2}}{4\kappa(1+\xi^{2})^{2}}, \label{solB}
\end{eqnarray}
which is exactly the expression (\ref{bpb1}). We finalize by noting that, unfortunately, finding the analytic solution of Eq.~(\ref{two_B}) or 
Eq.~(\ref{pnn2}), incorporating both single- and two-photon processes, is not a trivial task.


\end{document}